\documentclass[11pt]{article}

\usepackage{amsmath}
\usepackage{amssymb}

\usepackage{graphicx}

\usepackage{cite}

\usepackage{color}

\usepackage{setspace}
\usepackage[labelfont=bf,labelsep=period,justification=raggedright]{caption}

\makeatletter
\renewcommand{\@biblabel}[1]{\quad#1.}
\makeatother

\date{}

\usepackage{url}
\usepackage{longtable}
\usepackage{fixltx2e}
\begin{document}

\begin{spacing}{1.2}
% Title must be 150 characters or less
\begin{flushleft}
{\Large
\textbf{ss3sim: An \textsf{R} package for fisheries stock assessment simulation with Stock Synthesis}
}
% Insert Author names, affiliations and corresponding author email.
\\
Sean C. Anderson$^{1,\ast}$,
Cole C. Monnahan$^{2}$,
Kelli F. Johnson$^{3}$,
Kotaro Ono$^{3}$,
Juan L. Valero$^{4}$
\\
\bf{1} Department of Biological Sciences,
Simon Fraser University,
Burnaby BC, V5A 1S6, Canada
\\
\bf{2} Quantitative Ecology and Resource Management,
University of Washington, Box 352182,
Seattle, WA 98195-2182, USA
\\
\bf{3} School of Aquatic and Fishery Sciences,
University of Washington, Box 355020,
Seattle, WA 98195-5020, USA
\\
\bf{4} Center for the Advancement of Population Assessment Methodology
(CAPAM), 8901 La Jolla Shores Drive, La Jolla, CA 92037, USA
\\
$\ast$ E-mail: sean@seananderson.ca
\end{flushleft}

%\linenumbers
%\modulolinenumbers[1]

%\input{manuscript-plos.tex}
\section*{Abstract}

Simulation testing is an important approach to evaluating fishery stock
assessment methods. In the last decade, the fisheries stock assessment modeling
framework Stock Synthesis (SS3) has become widely used around the world.
However, there lacks a generalized and scriptable framework for SS3 simulation
testing. Here, we introduce \textbf{ss3sim}, an \textsf{R} package that
facilitates reproducible, flexible, and rapid end-to-end simulation
testing with SS3. \textbf{ss3sim} requires an existing SS3 model configuration
along with plain-text control files describing alternative population dynamics,
fishery properties, sampling scenarios, and assessment approaches.
\textbf{ss3sim} then generates an underlying `truth' from a specified
operating model, samples from that truth, modifies and runs an estimation
model, and synthesizes the results. The simulations can be run in parallel,
reducing runtime, and the source code is free to be modified under an
open-source MIT license. \textbf{ss3sim} is designed to explore structural
differences between the underlying truth and assumptions of an estimation
model, or between multiple estimation model configurations. For example,
\textbf{ss3sim} can be used to answer questions about model misspecification,
retrospective patterns, and the relative importance of different types of
fisheries data. We demonstrate the software with an example, discuss how
\textbf{ss3sim} complements other simulation software, and outline specific
research questions that \textbf{ss3sim} could address.

\clearpage
\section*{Introduction}

Fisheries stock assessment models are an invaluable tool for providing
scientific advice regarding stock status, historical productivity, and
changes in stock composition as well as evaluating the impact of alternative
management actions on fishery resources \cite{gulland1983, hilborn1992}.
Although a variety of stock assessment approaches are available, it is often
not straightforward to select among competing alternatives that may lead
to different conclusions about stock status and associated scientific
advice to management.

Simulation testing is a critical component to understanding the behavior of
fishery stock assessment methods, particularly given the potential for model
misspecification \cite{hilborn1987, hilborn1992, rosenberg1994, peterman2004,
  deroba2014}. With simulation testing we can evaluate the precision and bias
of alternative assessment approaches in a controlled environment where we know
the true dynamics of hypothetical fisheries resources under exploitation.
Recent simulation studies have been key to improving structural
assumptions for dealing with, for example, time-varying natural mortality ($M$)
\cite{lee2011, jiao2012, deroba2013, johnson2014}, uncertainty in
steepness of the stock-recruit relationship \cite{lee2012}, and environmental
variability \cite{schirripa2009}, as well as determining the utility and
influence on assessment outcomes of various fishery-dependent and -independent
data sources \cite{magnusson2007, wetzel2011a, ono2014, yin2004}.

There is a suite of tools available for conducting fishery stock
assessments and Stock Synthesis (SS3, the third version of the software) is
one, widely-used, modeling framework \cite{methot2013}. SS3 implements
statistical age-structured population modeling using a wide range of minimally
processed data \cite{maunder2013, methot2013}. The generalized model
structure of SS3 allows flexible scaling to a variety of data and life-history
situations, from data-poor (e.g.~\cite{wetzel2011a, cope2013}) to
data-rich (e.g.~\cite{haltuch2013}). Owing in part to these advantages,
SS3 has been used worldwide to formally assess 61 fishery stocks by 2012:
35 stocks in the US, 10 tuna/billfish stocks in three oceans, four European
stocks, and 12 Australian stocks \cite{methot2013}. These assessments are
conducted by both national agencies (e.g.~NOAA in the USA, CSIRO in Australia)
as well as regional fisheries management organizations (e.g.~IATTC, ICCAT, IOTC
in the Pacific, Atlantic and Indian oceans respectively). In addition to
completed formal stock assessments, exploratory SS3 applications for many other
stocks are underway \cite{methot2013}.

Stock Synthesis is also commonly used as a framework for stock assessment
simulation testing \cite{helu2000, yin2004, schirripa2009, lee2011, jiao2012,
  lee2012, crone2013a, hurtadoferro2013}, but there lacks a generalized
structure for simulation testing with SS3. As a result, most stock
assessment simulation-testing work using SS3 to date has relied on custom
frameworks \cite{helu2000, yin2004, magnusson2007, wetzel2011a, jiao2012,
  wilberg2006, deroba2013, deroba2014, crone2013a, hurtadoferro2013}.
Although custom-designed modeling frameworks can be tailored to the
specific needs of a particular stock assessment or simulation study, the use of
a generalized framework allows other scientists to validate, reuse, and build
upon previous work, thereby improving efficiency and resulting in more reliable
outcomes.

The programming language \textsf{R} \cite{rcoreteam2013} is an ideal language
in which to write such a generalized framework because (1) \textsf{R} has
become the standard for statistical computing and visualization and (2) the
\textsf{R} package \textbf{r4ss} \cite{r4ss2013} facilitates reading,
processing, and plotting of SS3 model output. Here we introduce
\textbf{ss3sim}, an \textsf{R} package that facilitates reproducible,
flexible, and rapid end-to-end simulation testing with the widely-used SS3
framework. We begin by outlining the general structure of \textbf{ss3sim} and
describing its functions, and then demonstrate the software with a simple
example. We conclude by discussing how \textbf{ss3sim} complements other
simulation testing software and by outlining some research questions that our
freely accessible and general simulation-testing framework could address.

\section*{The ss3sim framework}

\subsection*{Design goals of ss3sim}

We designed \textbf{ss3sim} simulations to be reproducible,
flexible, and rapid. \emph{Reproducible}: \textbf{ss3sim} simulations are
produced using \textsf{R} code, plain-text control files, and SS3 model
configurations. \textbf{ss3sim} also allows for random seeds to be set when
generating observation and process error. In combination, these features make
simulations repeatable across computers and operating systems (Windows, OS X,
and Linux). \emph{Flexible}: \textbf{ss3sim} inherits the flexibility of SS3
and can therefore implement many available stock assessment configurations by
either modifying existing SS3 model configurations or by modifying generic
life-history model configurations that are built into \textbf{ss3sim} (Text
S1). Furthermore, \textbf{ss3sim} summarizes the simulation output into
plain-text comma-separated-value (\texttt{.csv}) files allowing the output to
be processed in \textsf{R} or other statistical software. Finally, the
\textbf{ss3sim} source code is written under an open-source MIT license and can
be freely modified. \emph{Rapid}: \textbf{ss3sim} relies on SS3, which uses AD
Model Builder \cite{fournier2012} --- a rapid and robust
non-linear optimization software \cite{bolker2013} --- as a back-end
optimization platform. \textbf{ss3sim} also facilitates the deployment of
simulations across multiple computers or computer cores (i.e.~parallelization),
thereby reducing runtime. By using the vetted SS3 framework with the
tested \textbf{ss3sim} package, the time to develop and run a large-scale
simulation study can be reduced substantially, allowing for more time to
refine research questions and interpret results instead of spending it
developing and testing custom simulation frameworks.

\subsection*{The general structure of an ss3sim simulation}

\textbf{ss3sim} consists of both low-level functions that modify SS3
configuration files and high-level functions that combine these low-level
functions into a complete simulation experiment (Figure 1, Table 1). In this
paper we will focus on the structure and use of the high-level function
\texttt{run\_ss3sim}; however, the low-level functions can be used on their own
as part of a customized simulation (see Text S1).

An \textbf{ss3sim} simulation requires three types of input: (1) a base SS3
model configuration describing the underlying true population dynamics, or
operating model (OM); (2) a base SS3 model configuration to assess the observed
data generated by the OM, also known as the estimation model or method (EM);
and (3) a set of plain-text files (case files) describing alternative model
configurations and deviations from these base models (e.g.~different fishing
mortality or $M$ trajectories; Figure 2). We refer to each unique
combination of OM, EM, and case files as a scenario. Scenarios are usually run
for multiple iterations with unique process and observation error in each
iteration. An \textbf{ss3sim} simulation therefore refers to the combination of
all scenarios and iterations.

The \texttt{run\_ss3sim} function works by modifying SS3 configuration files as
specified in the case-file arguments (\texttt{change} functions), running the
OM, sampling from the time-series of true population dynamics to generate an
observed dataset (\texttt{sample} functions), running the EM to get
maximum-likelihood estimates and standard errors of parameters and
derived quantities, and synthesizing the output for easy data
manipulation and visualization (\texttt{get} functions) (Figure 1).

\section*{An example simulation with ss3sim}

To demonstrate \textbf{ss3sim}, we will work through a simple example in which
we examine the effect of (1) high vs.~low precision of a fishery independent
index of abundance and (2) fixing $M$ at an assumed value vs.~estimating
$M$. All files to run this example are included in the package data, and a more
detailed description is available in the accompanying vignette (Text S1).
\textbf{ss3sim} requires \textsf{R} version 3.0.0 or greater and SS3 (see Text
S1 for more detailed instructions). In \textsf{R}, \textbf{ss3sim} can be
installed and loaded with:

\begin{verbatim}
install.packages("ss3sim")
library("ss3sim")
\end{verbatim}

\noindent
Alternatively, the development version of \textbf{ss3sim} can be installed from
\url{https://github.com/ss3sim/ss3sim}. You can read the documentation and
vignette (Text S1) with:

\begin{verbatim}
?ss3sim
vignette("ss3sim-vignette")
\end{verbatim}

\subsection*{Setting up the SS3 model configurations}

\textbf{ss3sim} comes with built-in SS3 model configurations that represent
three general life histories: cod-like (slow-growing and long-lived),
flatfish-like (fast-growing and long-lived), and sardine-like (fast-growing and
short-lived). These model configurations are based on North Sea cod
(\emph{Gadus morhua}; R. Methot, NMFS, NOAA; pers.~comm.), yellowtail flounder
(\emph{Limanda ferruginea}; R. Methot, NMFS, NOAA; pers. comm.), and Pacific
sardine (\emph{Sardinops sagax caeruleus}) \cite{hill2012} (Text S1). We
recommend modifying these built-in model configurations to match a desired
scenario, although it is possible to create a new \textbf{ss3sim} model by
modifying an existing SS3 model configuration (Text S1). We will base our
example around the built-in cod-like model setup.

\subsection*{Setting up the case files}

The high-level function \texttt{run\_ss3sim} can run all simulation steps based
on a specified scenario ID and a set of semicolon-delimited plain-text files
that describe alternative cases (Figures 1 and 2). These case files
contain argument values that are passed to the low-level \textbf{ss3sim}
\textsf{R} functions (e.g.~\texttt{change\_index}, a function that controls how
the fishery and survey indices are sampled; Table 1).

To use \texttt{run\_ss3sim}, all case files must be named according to the type
of case (e.g.~\texttt{E} for estimation or \texttt{F} for fishing mortality), a
numeric value representing the case number, and an alphanumeric identifier
representing the species or stock (e.g.~\texttt{cod}; Table 1, Text S1). We
combine these case IDs with hyphens to create scenario IDs. For example, one of
our scenarios will have the scenario ID \texttt{D1-E0-F0-M0-R0-cod}. This
scenario ID tells \texttt{run\_ss3sim} to read the case files corresponding to
the first data (\texttt{D}) case (i.e.~\texttt{index1-cod.txt},
\texttt{lcomp1-cod.txt}, \texttt{agecomp1-cod.txt)}, the zero case for
estimation (\texttt{E}; i.e.~\texttt{E0-cod.txt)}, and so on.

To investigate the effect of different levels of precision of a
fishery-independent index of abundance, we will manipulate the argument
\texttt{sds\_obs} that gets passed to the function \texttt{change\_index}. In
data case \texttt{D0}, we will specify the standard deviation of the index of
abundance at 0.1 and in case \texttt{D1} we will increase the standard
deviation to 0.4. We can do this by including the line: \texttt{sds\_obs;
  list(0.1)} in the file \texttt{index0-cod.txt} and the line:
\texttt{sds\_obs; list(0.4)} in the file \texttt{index1-cod.txt}. We will also
set up a base-case file describing fishing mortality (\texttt{F0-cod.txt}), a
file describing a stationary $M$ trajectory (\texttt{M0-cod.txt}), and specify
that we do not want to run a retrospective analysis in the file
\texttt{R0-cod.txt}. We will set up the file \texttt{E0-cod.txt} to fix $M$ at
the true value and not estimate it, and case \texttt{E1-cod.txt} to estimate a
stationary, time-invariant $M$ (Text S1).

All of these text files are available in the package data in the folder
\texttt{inst/extdata/eg-cases/}. As an example, here is what the complete
\texttt{index0-cod.txt} file looks like:

\begin{verbatim}
fleets; 2
years; list(seq(1974, 2012, by = 2))
sds_obs; list(0.1)
\end{verbatim}

\noindent
\texttt{fleets}, \texttt{years}, and \texttt{sds\_obs} refer to the arguments
in the function \texttt{change\_index} and users can read the help for this
function with \texttt{?change\_index} in \textsf{R}.

\subsection*{Validating the simulation setup}

Before running and interpreting the results of a simulation, it is
important to validate the testing framework at several levels. First, it is
important to test that the functions that manipulate model configurations
(i.e.~the \texttt{change} functions) are set up properly. \textbf{ss3sim} comes
with prepackaged models that have been tested extensively with the
\texttt{change} functions, as well as documented \textsf{R} functions that
include examples and unit tests. We describe strategies for testing the
\texttt{change} functions on new SS3 model setups in Text S1.

Second, the components of the simulation framework must work together as
expected (integration tests \cite{wilson2014}). One approach to testing for
such issues is to run simulation tests with similar OM and EM setups and
relatively low process and observation error \cite{hilborn1992}.
\textbf{ss3sim} makes this form of validation simple by allowing users to
specify levels of process and observation error (Text S1). Assuming
that the user specifies sufficient error to avoid numerical instability, this
approach can reveal issues that would otherwise be obscured by noise.

Finally, it is important to validate that the model-fitting algorithms
converged to global maxima. \textbf{ss3sim} retains all SS3 model output for
future examination, as well as performance diagnostics such as maximum
gradient, whether or not the covariance matrix was successfully calculated, run
time, and the number of parameters stuck on bounds. These metrics, in
combination with visual checks, are useful to determine if the results of a
study are robust and meaningful.

\subsection*{Running the simulations}

Since we have already validated the cod-like model setup (Text S1), we can now
run our example simulation scenario. To start, we will locate three sets of
folders within the package data: the folder with the OM, the folder with the
EM, and the folder with the plain-text case files:

\begin{verbatim}
d <- system.file("extdata", package = "ss3sim")
om <- paste0(d, "/models/cod-om")
em <- paste0(d, "/models/cod-em")
case_folder <- paste0(d, "/eg-cases")
\end{verbatim}

We can then run the simulation with one call to the \texttt{run\_ss3sim}
function. We will set \texttt{bias\_adjust = TRUE} to enable a procedure that
aims to produce mean-unbiased estimates of recruitment and biomass despite
log-normal recruitment deviations \cite{methot2011}. We can run 100 iterations
of the simulation scenarios with the following code:

\begin{verbatim}
run_ss3sim(iterations = 1:100, scenarios =
  c("D0-E0-F0-M0-R0-cod", "D1-E0-F0-M0-R0-cod",
    "D0-E1-F0-M0-R0-cod", "D1-E1-F0-M0-R0-cod"),
  case_folder = case_folder, om_dir = om,
  em_dir = em, bias_adjust = TRUE)
\end{verbatim}

\noindent
This produces a folder structure in our working directory containing all of the
SS3 output files (Figure 2). We can then collect the output with one
function call:

\begin{verbatim}
get_results_all()
\end{verbatim}

\noindent
This command creates two files in our working directory:
\texttt{ss3sim\_scalars.csv} and \texttt{ss3sim\_ts.csv}, which contain scalar
output estimates (model parameters and derived quantities such as
steepness and maximum sustainable yield) and time-series estimates
(e.g.~recruitment and biomass for each year). These estimates come from the
report files produced from each run of SS3 as extracted by the \textbf{r4ss}
\textsf{R} package. The \texttt{.csv} files contain separate columns for OM and
EM values, making it simple to calculate error metrics, such as relative or
absolute error. In addition to parameter estimates, the \texttt{.csv} files
contain performance metrics, which in combination can be used to gauge model
performance and convergence. These results are organized into ``long'' data
format, with columns for scenario and iteration, facilitating quick analysis
and plotting using common \textsf{R} packages such as \textbf{ggplot2}
\cite{wickham2009}.

For the example simulation, the relative error in spawning stock biomass over
time is, as expected, smaller when the true value of $M$ is specified rather
than estimated (Figure 3, top panels E0 vs.~E1). Furthermore, lower precision
in the research survey index of abundance results in greater relative error in
spawning stock biomass in recent years (Figure 3, top panels D0 vs.~D1), and
greater relative error in terminal-year depletion (the ratio of terminal year
spawning biomass to unfished spawning biomass) and fishing mortality, but not
in spawning stock biomass at maximum sustainable yield, or $M$ (Figure 3, lower
panels).

\section*{How ss3sim complements other simulation software}

The general purpose of \textbf{ss3sim} is to explore model behaviour and
performance across combinations of EM configurations and alternative dynamics
of fisheries resources under exploitation specified by the OM. In particular,
\textbf{ss3sim} provides a suite of functions for dynamically creating
structural differences in both OMs and EMs. This expedites testing the
properties of alternative stock assessment model configurations, whether the
differences are between OMs and EMs \cite{johnson2014}, or between multiple
versions of EMs \cite{ono2014}. However, \textbf{ss3sim} is less suited for
quickly exploring new SS3 model setups, which may rely on SS3 configurations
not yet converted to work with the \textbf{ss3sim} package functions.
Although it is possible to adapt arbitrary SS3 models to work with
\textbf{ss3sim} (Text S1), other software frameworks may provide better
alternatives, depending on the goal of the simulation study.

One alternative software framework is Fisheries Libraries in \textsf{R} (FLR)
\cite{kell2007} --- a collection of open-source \textsf{R} packages developed
specifically for evaluating fisheries management strategies through simulation.
Compared to \textbf{ss3sim}, FLR is designed to explore broader questions
regarding management strategies with flexible biological, economic, and
management components \cite{hillary2009}. Thus, it is not specifically designed
to explore the impact of structural differences within OMs and EMs.

Another alternative stock assessment simulation testing framework is Fishery
Simulation (FS, \url{http://fisherysimulation.codeplex.com}). FS is primarily a
file management tool adapted to aid in simulation testing. FS can work with
stock assessment models besides SS3, make simple changes to input text files,
generate simple random process errors (using a built-in random number
generator) and observation errors (using the SS3 bootstrap option), run
simulations in parallel, and collect results from output files. Thus, FS is
closer to \textbf{ss3sim} in its scope than FLR in that it specifically focuses
on the performance of stock assessment models. FS differs from \textbf{ss3sim}
mainly in that it uses user-specified text manipulation commands (e.g.~change
line 50 from 0 to 1) to alter model configurations rather than the approach of
\textbf{ss3sim}, which uses modular functions tailored to specific purposes
(e.g.~add a particular time-varying mortality trajectory to an arbitrary OM).
FS works well for testing arbitrary assessment models and model configurations
because it does not rely on pre-built manipulation functions \cite{lee2012,
  piner2011, lee2011}. In contrast, FS cannot make complicated structural
changes to a model setup (e.g.~adding time-varying parameters or changing the
survey years), limiting its ability to induce and test structural differences
between OMs and EMs. In addition, the current version of FS is not an
end-to-end package --- additional code is necessary to incorporate arbitrary
process and observation error in simulation testing. Finally, although FS is
also open-source, it requires the Microsoft .NET framework and is therefore
only compatible with the Windows operating system.

\section*{Research opportunities with ss3sim}

The \textbf{ss3sim} package has been used so far to evaluate alternative
assessment approaches when $M$ is thought to vary across time
\cite{johnson2014}, the effect of various qualities and quantities of
length- and age-composition data on the bias and accuracy of assessment model
estimates \cite{ono2014}, and the causes of retrospective patterns in stock
assessment model estimates. Along with those studies, \textbf{ss3sim} makes
many relevant research opportunities easily approachable. Below we outline some
examples.

\emph{Time-varying model misspecification}: Ecological processes can vary
through time in response to, for example, changes to fishing behaviour
\cite{hilborn1992}, regime shifts \cite{vert-pre2013}, or climate change
\cite{walther2002}. However, parameters such as $M$, catchability, and
selectivity are commonly assumed to be time invariant and the consequences of
these assumptions when facing true temporal changes has been a long-standing
discussion in fisheries science \cite{royama1992, wilberg2006, fu2001}.
Furthermore, although studies have tried to isolate the effects of single
time-varying parameter, such as $M$ \cite{lee2011, jiao2012, deroba2013,
  johnson2014}, few have considered the effects of multiple time-varying
parameters and their potential interaction. \textbf{ss3sim} can easily turn
parameter estimation on and off as well as add time-varying dynamics to the OM,
making it an ideal candidate for assessing the effects of multiple time-varying
parameters.

\emph{Patterns in recruitment deviations}: Typically, estimation methods assume
independent log-normally-distributed recruitment deviations around a spawning
stock recruitment function. However, recruitment deviations are frequently
auto-correlated and their variability can change through time
\cite{beamish1995, pyper1998}. \textbf{ss3sim} facilitates exploring the
effect of different recruitment deviation structures on model performance by
allowing the user to directly specify any vector of deviations.

\emph{Retrospective patterns}: Retrospective patterns, in which model estimates
are systematically biased with each additional year of data, are a major
problem in stock assessment science \cite{mohn1999, legault2008}. Key questions
include what causes retrospective patterns and which assessment approaches reduce
them \cite{legault2008}. \textbf{ss3sim} can run retrospective analyses as part
of any simulation by adding a single argument: the number of retrospective
years to investigate.

\section*{Conclusions}

The increasing complexity of modern integrated stock assessment models and
expanding computing power allows for the inclusion of multiple sources of data
and estimation of increasingly complex processes \cite{maunder2013}. However,
it is difficult to determine under which conditions these processes can be
reliably estimated based on diagnostics such as residual patterns
\cite{maunder2013}. Simulation testing is an important tool because it provides
an opportunity to explore model performance under specified conditions and
develop a further understanding of a model's abilities.

We anticipate that \textbf{ss3sim} will facilitate the development of reliable
assessment methods, applicable to age-structured stock assessment frameworks in
general, that meet the requirements and assessment demands of many regional
fisheries management organizations and national assessment agencies. For
example, Johnson et~al.~\cite{johnson2014} used \textbf{ss3sim} to
develop guidelines for how to model natural mortality (when it is suspected of
being time varying but age invariant) across life histories and fishing
patterns. As another example, Ono et~al.~\cite{ono2014} used
\textbf{ss3sim} to identify the most informative combination of quantity,
quality, and timing of data, depending on life history and
stock-assessment-derived metrics of interest. General guidelines such as these,
combined with simulations testing specific model configurations used by
assessment agencies, are an important part of developing reliable assessment
methods to provide sound scientific advice to fisheries management
\cite{deroba2014, crone2013}.

Custom-tailored simulation-testing software packages are an increasingly common
tool in fisheries science, but their value would be extended if shared formally
with the broader community. Published, open-source simulation frameworks, such
as the initial release of \textbf{ss3sim} described here, allow other
scientists to validate, reuse, and improve the software. We therefore encourage
authors to publish their simulation frameworks and develop them in a
generalized format, where possible. We anticipate that users will both benefit
from \textbf{ss3sim} in its current form and extend it for their own needs,
potentially contributing to future versions.

\section*{Acknowledgements}

We thank the participants and mentors of the University of Washington's School
of Aquatic and Fishery Sciences 2013 FISH 600 course. Discussions with these
individuals were instrumental to the conceptual and technical development of
\textbf{ss3sim}. Many participants also contributed code and are listed within
specific \textbf{ss3sim} \textsf{R} functions. Participants: Curry Cunningham,
Felipe Hurtado-Ferro, Roberto Licandeo, Carey McGilliard, Melissa Muradian,
Cody Szuwalski, Katyana Vert-pre, and Athol Whitten. Mentors: Richard Methot,
Andr\'{e} Punt, Jim Ianelli, and Ian Taylor. We thank Jim Ianelli, Andr\'{e}
Punt, Ian Stewart, Robert Ahrens, Daniel Duplisea, and an anonymous reviewer
for comments that greatly improved our manuscript.

SCA was supported by Fulbright Canada (generously hosted by Trevor Branch),
NSERC, and a Garfield Weston Foundation/B.C. Packers Ltd. Graduate Fellowship
in Marine Sciences. This work was partially funded by the Joint Institute for
the Study of the Atmosphere and Ocean (JISAO) under NOAA Cooperative Agreement
No.~NA10OAR4320148, Contribution No.~2193. This research addresses the methods
component of the good practices guide to stock assessment program of the Center
for the Advancement of Population Assessment Methodology (CAPAM).

%\bibliography{ss3sim-ms}

\clearpage

\section*{Figure Legends}

\begin{figure}[!ht]
 \begin{center}
 \includegraphics[width=3.27in]{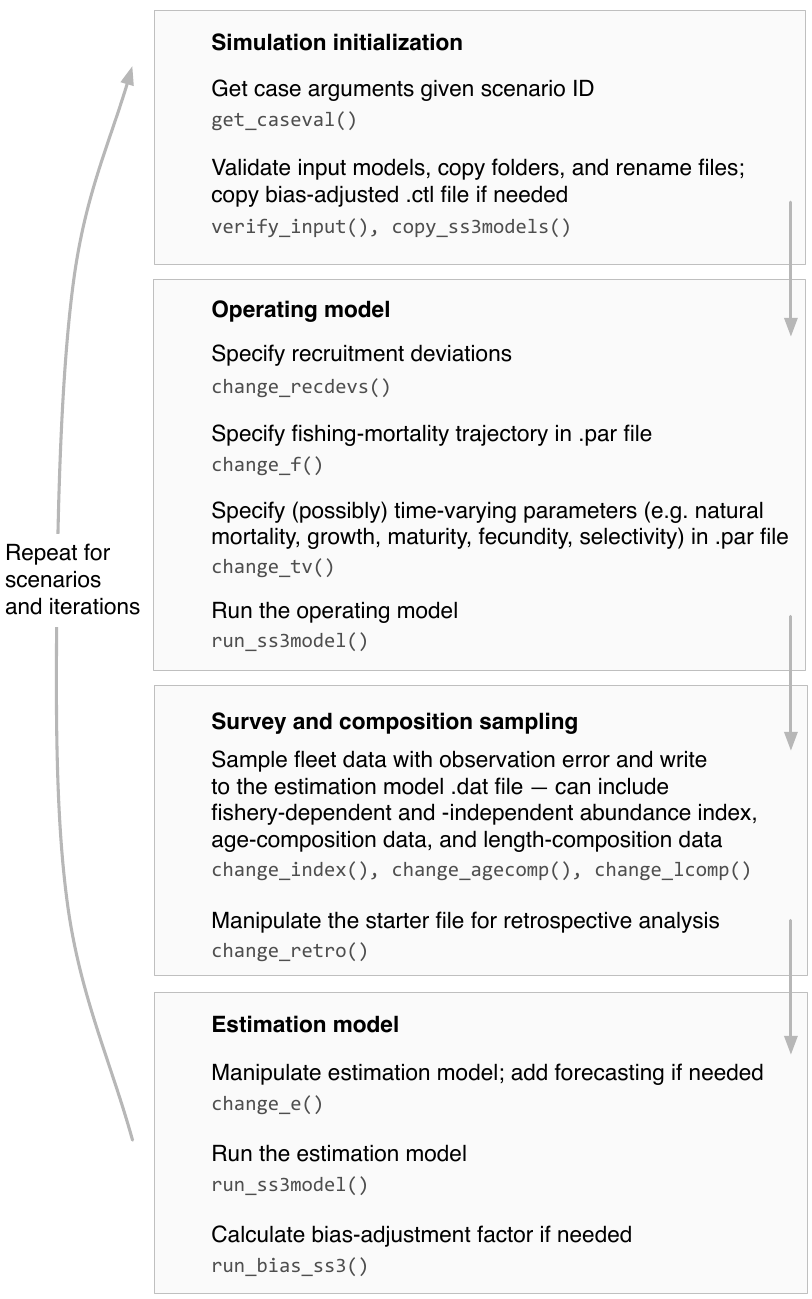}
 \end{center}
\caption{
{\bf Flow diagram of the main steps in an ss3sim simulation carried out using
  \texttt{run\_ss3sim}.} Functions that are called internally are shown in a
monospaced font.
}
\label{fig:sim-steps}
\end{figure}

\clearpage

\begin{figure}[!ht]
 \begin{center}
 \includegraphics[width=4.86in]{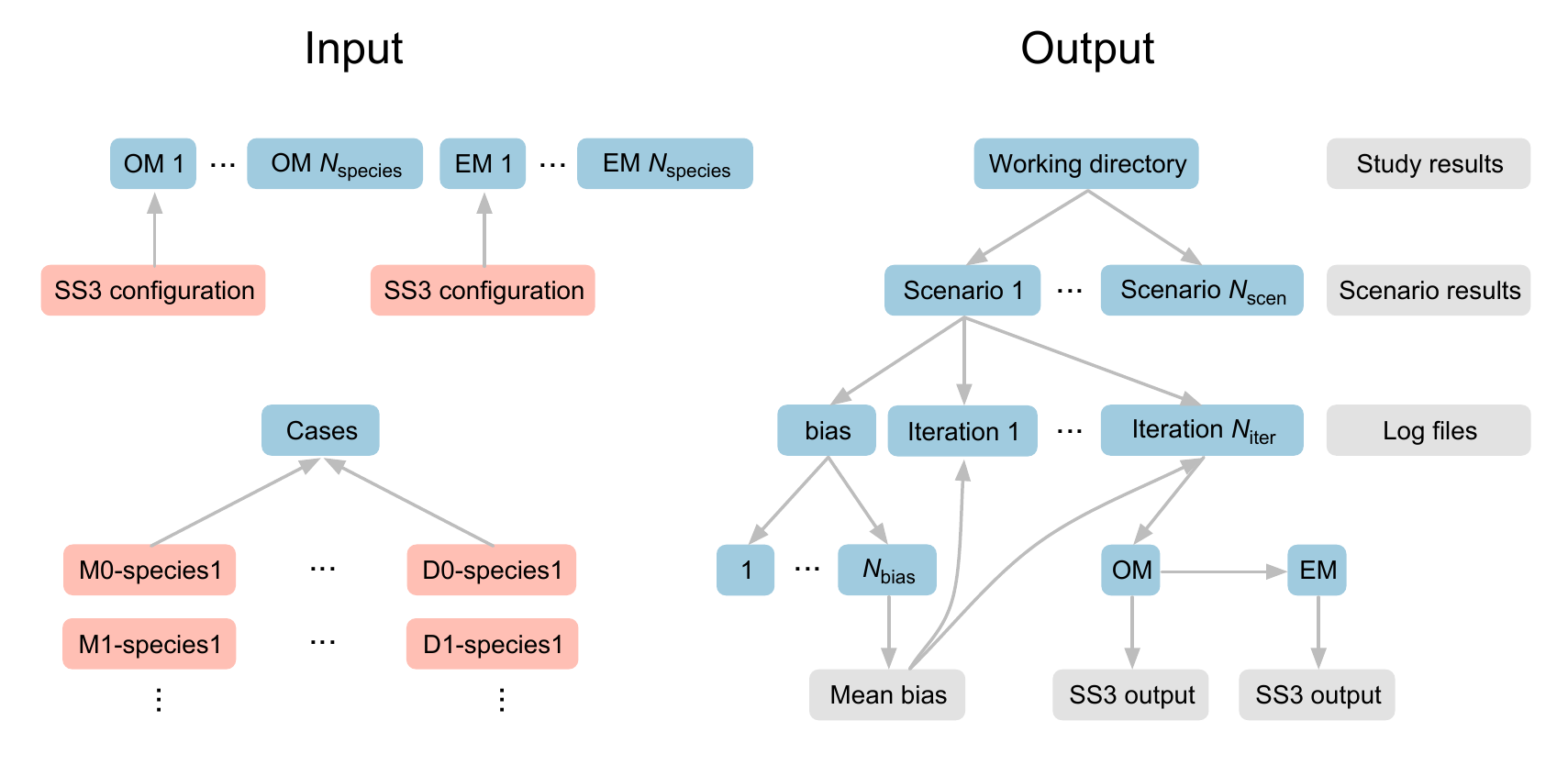}
 \end{center}
\caption{
{\bf Illustration of input and output folder and file structure for an ss3sim
  simulation.} Folders are shown in blue, input files in orange, and output
files in grey. All input and output files are in plain text format. OM refers
to operating model and EM to estimation model. Case files (orange files at
bottom left) combine cases (e.g.~\texttt{M0} for a given natural mortality
trajectory) with species or life-history OMs and EMs (e.g.~cod-like or
sardine-like). Alternatively, a user can skip setting up case files and specify
the simulation cases directly in \textsf{R} code (see the accompanying vignette
[Text S1]).
}
\end{figure}

\clearpage

\begin{figure}[!ht]
 \begin{center}
 \includegraphics[width=4.86in]{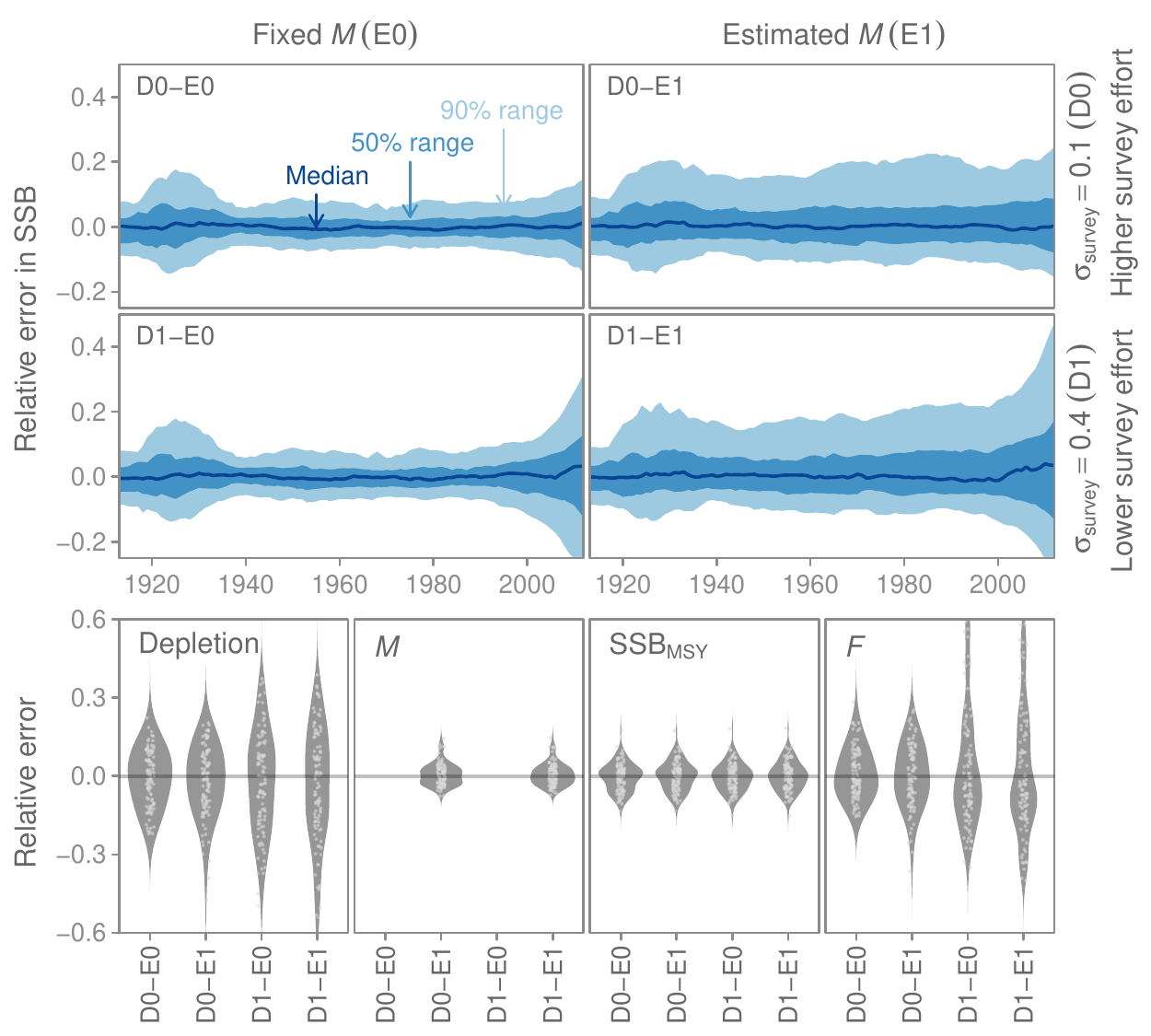}
 \end{center}
\caption{
{\bf Example output from an ss3sim simulation.} We ran a crossed simulation in
which we considered (1) the effect of fixing natural mortality ($M$) at its
true value (0.2; case E0) or estimating $M$ (case E1) and (2) the effect of
high survey effort ($\sigma_\mathrm{survey} = 0.1$; case D0) or low survey
effort ($\sigma_\mathrm{survey} = 0.4$; case D1). Upper panels (blue) show time
series of relative error in spawning stock biomass (SSB). Lower panels (grey)
show the distribution of relative error across four scalar variables: depletion
(the ratio of terminal year spawning biomass to unfished spawning biomass),
$M$, SSB at maximum sustainable yield ($\mathrm{SSB}_\mathrm{MSY}$), and
fishing mortality ($F$) in the terminal year. We show the values across
simulation iterations with dots and the distributions with beanplots (kernel
density smoothers).
}
\label{fig:results}
\end{figure}

\clearpage

\section*{Tables}

\textbf{Table 1. Main ss3sim functions and a description of their purpose.}
Simulations can be run through the \texttt{run\_ss3sim} function, which then
calls the \texttt{change} functions. Users can control what the \texttt{change}
functions do through a series of plain-text case files. For example, the case
ID \texttt{D1} corresponds to the case files \texttt{lcomp1},
\texttt{agecomp1}, and \texttt{index1}, as described in the table. Users can
also skip setting up case files and specify arguments to \texttt{ss3sim\_base}
directly, or use the \texttt{change} functions as part of their own simulation
structure (Text S1).

\begin{longtable}[c]{@{}ll@{}}
\hline\noalign{\medskip}
\begin{minipage}[b]{0.32\columnwidth}\raggedright
Function name
\end{minipage} & \begin{minipage}[b]{0.57\columnwidth}\raggedright
Description
\end{minipage}
\\\noalign{\medskip}
\hline\noalign{\medskip}
\begin{minipage}[t]{0.32\columnwidth}\raggedright
\texttt{run\_ss3sim}
\end{minipage} & \begin{minipage}[t]{0.57\columnwidth}\raggedright
Main high-level function to run \textbf{ss3sim} simulations.
\end{minipage}
\\\noalign{\medskip}
\begin{minipage}[t]{0.32\columnwidth}\raggedright
\texttt{ss3sim\_base}
\end{minipage} & \begin{minipage}[t]{0.57\columnwidth}\raggedright
Underlying base simulation function. Can also be called directly.
\end{minipage}
\\\noalign{\medskip}
\begin{minipage}[t]{0.32\columnwidth}\raggedright
\texttt{change\_rec\_devs}
\end{minipage} & \begin{minipage}[t]{0.57\columnwidth}\raggedright
Substitutes recruitment deviations.
\end{minipage}
\\\noalign{\medskip}
\begin{minipage}[t]{0.32\columnwidth}\raggedright
\texttt{change\_f}
\end{minipage} & \begin{minipage}[t]{0.57\columnwidth}\raggedright
Adds fishing mortality time series. (Case file and ID \texttt{F})
\end{minipage}
\\\noalign{\medskip}
\begin{minipage}[t]{0.32\columnwidth}\raggedright
\texttt{change\_tv}
\end{minipage} & \begin{minipage}[t]{0.57\columnwidth}\raggedright
Adds time-varying features. For example, time-varying natural mortality,
growth, or selectivity. (Any case file and ID, e.g.~\texttt{M}, starting with
``\texttt{function\_type; change\_tv}'')
\end{minipage}
\\\noalign{\medskip}
\begin{minipage}[t]{0.32\columnwidth}\raggedright
\texttt{change\_index}
\end{minipage} & \begin{minipage}[t]{0.57\columnwidth}\raggedright
Controls how the fishery and survey indices are sampled. (Case file
\texttt{index}, case ID \texttt{D})
\end{minipage}
\\\noalign{\medskip}
\begin{minipage}[t]{0.32\columnwidth}\raggedright
\texttt{change\_agecomp}
\end{minipage} & \begin{minipage}[t]{0.57\columnwidth}\raggedright
Controls how age composition data are sampled. (Case file \texttt{agecomp},
case ID \texttt{D})
\end{minipage}
\\\noalign{\medskip}
\begin{minipage}[t]{0.32\columnwidth}\raggedright
\texttt{change\_lcomp}
\end{minipage} & \begin{minipage}[t]{0.57\columnwidth}\raggedright
Controls how length composition data are sampled. (Case file \texttt{lcomp},
case ID \texttt{D})
\end{minipage}
\\\noalign{\medskip}
\begin{minipage}[t]{0.32\columnwidth}\raggedright
\texttt{change\_retro}
\end{minipage} & \begin{minipage}[t]{0.57\columnwidth}\raggedright
Controls the number of years to discard for a retrospective analysis. (Case
file and ID \textsf{R})
\end{minipage}
\\\noalign{\medskip}
\begin{minipage}[t]{0.32\columnwidth}\raggedright
\texttt{change\_e}
\end{minipage} & \begin{minipage}[t]{0.57\columnwidth}\raggedright
Controls which and how parameters are estimated. (Case file and ID \texttt{E})
\end{minipage}
\\\noalign{\medskip}
\begin{minipage}[t]{0.32\columnwidth}\raggedright
\texttt{run\_bias\_ss3}
\end{minipage} & \begin{minipage}[t]{0.57\columnwidth}\raggedright
Determines the level of adjustment to ensure mean-unbiased estimates of
recruitment and biomass.
\end{minipage}
\\\noalign{\medskip}
\begin{minipage}[t]{0.32\columnwidth}\raggedright
\texttt{get\_results\_scenario}
\end{minipage} & \begin{minipage}[t]{0.57\columnwidth}\raggedright
Extracts results for a single scenario.
\end{minipage}
\\\noalign{\medskip}
\begin{minipage}[t]{0.32\columnwidth}\raggedright
\texttt{get\_results\_all}
\end{minipage} & \begin{minipage}[t]{0.57\columnwidth}\raggedright
Extracts results for a series of scenarios.
\end{minipage}
\\\noalign{\medskip}
\hline
\end{longtable}

% \nolinenumbers
% \setlength\parskip{0.11in}
% \setlength\parindent{0in}
%
% \makeatletter
% \renewcommand\@seccntformat[1]{}
% \makeatother
%
% \clearpage
% \setcounter{page}{1}
%
% \onehalfspacing
% \input{ss3sim-response.tex}

\end{spacing}
\end{document}